\documentclass[12pt]{article}
\usepackage{amsmath,lscape}
\def\ii{\'{\i}}
\def\beq{\begin{equation}}
\def\eeq{\end{equation}}
\def\beq{\begin{equation}}
\def\eeq{\end{equation}}
\def\ban{\begin{eqnarray*}}
\def\ean{\end{eqnarray*}}
\def\bi{\begin{itemize}}
\def\ei{\end{itemize}}

\def\d{\mbox{d}}

\begin{document}

\begin{center}
{\bf NUCLEATION PROCESS IN ASYMMETRIC NUCLEAR MATTER}
\end{center}

\begin{center}
{\it D.P.Menezes $^*$ and C. Provid\^encia}\\
{\it Centro de F\ii sica Te\'orica - Depto de F\ii sica -
  Universidade de Coimbra}\\
{\it 3000 - Portugal}\\
{\it $^*$  Permanent Address: Depto de F\'{\i}sica - CFM -
  Universidade Federal de Santa Catarina}\\
{\it Florian\'opolis - SC - CP. 476 - CEP 88.040 - 900 - Brazil}
\end{center}

\vspace{0.50cm}

\begin{abstract}
An extended version of the non linear Walecka model, with $\rho$
mesons and eletromagnetic field is used to investigate the
possibility of phase transitions in hot (warm) nuclear matter,
giving rise to droplet formation. Surface properties of
asymmetric nuclear matter as the droplet surface energy and
its thickness are also examined.
\end{abstract}

\vspace{0.50cm}
PACS number(s):21.65.+f, 21.90.+f,26.60.+c

\newpage
                         
\section {Introduction}

In heavy-ion collisions, 
part of the hot nuclear 
matter produced can be described in terms of hadrons. The formation of 
highly excited 
composed nuclei in equilibrium with a gas of evaporated particles can be 
interpreted in the
framework of hydrodynamics as two coexisting phases of nuclear matter, 
a liquid and a gas
phase \cite{blv}. 
During these
reactions, phase transitions may occur depending on the temperature
and densities involved \cite{ms}-
\cite{bhrs}.
The investigation of 
asymmetric nuclear  matter  is also of particular interest for problems in
astrophysics. In fact, neutron-star matter at densities between 
$0.03$ fm$^{-3}$ and nuclear matter density ($0.17$ fm$^{-3}$) 
consists of neutron-rich nuclei immersed in a gas of neutrons
\cite{sha}. The size of the nuclei is determined by a competition between  
the surface energy 
and the Coulomb interaction.
The use of thermodynamical concepts
in the study of possible phase transitions in the above problems is done 
with the underlying assumption that the time required for
thermalization and chemical equilibrium is small.

Hot nuclei, liquid-gas phase transitions and droplet formation in nuclear 
reactions
as well as the surface properties of nuclear matter
have already been extensively discussed in the literature in the context of 
non--relativistic models, namely  within the
framework of  the Hartree-Fock (HF), Thomas--Fermi(TF) and
Extended--Thomas--Fermi approximations (ETF) \cite{blv},\cite{rpl}-
\cite{suraud}.
In particular, in \cite{suraud} it is shown that the semi-classical TF
approximation scheme
is reasonably accurate at any temperature. 

Within the framework of relativistic models 
phase transitions in hot (warm) nuclear matter have also been
investigated in infinite matter
by imposing constant mesonic fields \cite{ms},  
at zero temperature for symmetric semi-infinite  nuclear matter \cite{salzedas},
at finite temperature for symmetric matter in the linear Walecka model
\cite{rs},\cite{np} and also in its non--linear form for symmetric
and asymmetric matter \cite{md}.
Surface properties of asymmetric semi--infinite nuclear matter have
 been investigated at zero temperature in \cite{epsw} and
\cite{cev} in terms of a semi--classical
treatment.  In most of the above mentioned papers in which temperature
effects have been taken into account, the finite temperature
version of the liquid drop model is used in the investigation of
the surface properties of the arising droplet. 

In the present  work, we study the possibility of droplet formation in a 
vapour system at finite
temperature in the framework of  the relativistic  Walecka model with non 
linear terms
(NLWM), which is known to describe adequately  the bulk properties
of nuclear matter.  We include  the Coulomb interaction and work in the 
Thomas-Fermi approximation.
We determine the conditions for phase coexistence in a   multi-component system
by building the binodals for several temperatures and two parametrizations of 
the non-linear
Walecka model. These values determine the initial conditions which are 
used in solving 
numerically  the coupled equations of motion obtained 
 in  the Thomas-Fermi approximation at finite temperature. 
 This  semi-classical
approximation contains the following quantal
ingredients: the production of anti-particles is included and
the distribution function of particles and anti-particles takes into 
account the Pauli Principle.

In section 2 we obtain the equations of motion in the static case. The 
thermodinamical potential in the framework of the Thomas-Fermi
approximation
is calculated  in section 3. In section 4 the two--phase coexistence
is discussed and in  section 5
we present the numerical results. Finally,
in the last section some conclusions are drawn.

\section {Extended Non Linear Walecka Model}

In what follows we describe the equation of state of non-symmetric matter 
within the framework of the
relativistic non-linear Walecka model \cite{bb}, \cite{ring}
with the inclusion of 
$\rho$-mesons and the eletromagnetic field. 
The self interaction terms of the scalar meson were shown to be
necessary in order to adequately describe nuclear properties
\cite{bb}. Both the $\rho$ meson and photons are incorporated to account, 
respectively,
 for the neutron excess in heavy nuclei and the eletromagnetic interaction
between the protons \cite{ring}. In our lagrangian the $\pi$--meson field 
amplitude is not considered since we are not interested in pion--condensed
states and hence, under the approximation we use here, all pion
contributions vanish.

In this  model the nucleons are coupled to neutral scalar $\phi$, 
isoscalar-vector $V^\mu$, isovector-vector $\vec b^\mu$  meson fields and the
electromagnetic field $A^\mu$.
The lagrangian density reads:
$$
{\cal L}=\bar \psi\left[\gamma_\mu(i\partial^{\mu}-g_v V^{\mu})-
\frac{g_{\rho}}{2} \gamma_\mu \vec{\tau} \cdot \vec{b}^\mu 
-e \gamma_\mu A^\mu \frac {(1+ \tau_3)}{2} 
-(M-g_s \phi)\right]\psi
$$
$$
+\frac{1}{2}(\partial_{\mu}\phi\partial^{\mu}\phi
-m_s^2 \phi^2) - \frac{1}{3!}\kappa \phi^3 -\frac{1}{4!}\lambda
\phi^4
$$
$$
-\frac{1}{4}\Omega_{\mu\nu}\Omega^{\mu\nu}+\frac{1}{2}
m_v^2 V_{\mu}V^{\mu} + \frac{1}{4!}\xi g_v^4 (V_{\mu}V^{\mu})^2
$$
\begin{equation}
-\frac{1}{4}\vec B_{\mu\nu}\cdot\vec B^{\mu\nu}+\frac{1}{2}
m_\rho^2 \vec b_{\mu}\cdot \vec b^{\mu}  
-\frac{1}{4}F_{\mu\nu}F^{\mu\nu}\;,
\end{equation}
where
\begin{equation}
\Omega_{\mu\nu}=\partial_{\mu}V_{\nu}-\partial_{\nu}V_{\mu} ,
\end{equation}
\begin{equation}
\vec B_{\mu\nu}=\partial_{\mu}\vec b_{\nu}-\partial_{\nu} \vec b_{\mu}
- g_\rho (\vec b_\mu \times \vec b_\nu) ,
\end{equation}
and
\begin{equation}
F_{\mu\nu}=\partial_{\mu}A_{\nu}-\partial_{\nu}A_{\mu}.
\end{equation}
The model comprises the following parameters:
three coupling constants $g_s$, $g_v$ and $g_{\rho}$ of the mesons to
the nucleons, the nucleon mass $M$, the masses of
the mesons $m_s$, $m_v$, $m_{\rho}$, the
eletromagnetic coupling constant $e=\sqrt{\frac{4 \pi}{137}}$ and
the self-interacting coupling constants $\kappa$, $\lambda$ and $\xi$.
We have used two sets of constants, respectively identified as MS taken 
from \cite{ms} and as NL1 taken from 
\cite{sed}, with $C_i^2= g_i^2 M^2/m_i^2$, $i=~s,~v,~\rho,$
and they are displayed in table 1. 
All meson masses must
be specified along with the coupling constants. It is true that only the
rations of couplings to the masses are necessary in infinite matter in the
{\it linear} Walecka model. When non - linear terms are included, the
statement looses its validity.

From the Euler--Lagrange formalism, we obtain the coupled equations
of motion for the scalar, isoscalar-vector, isovector-vector,
eletromagnetic and nucleon fields, respectively given by:
\begin{equation}
(\partial_t^2-\nabla^2+m_s^2)\phi= g_s \rho_s - \frac{\kappa}{2} \phi^2
-\frac{\lambda}{6} \phi^3,
\end{equation}
\begin{equation}
(\partial_t^2-\nabla^2+m_v^2)V^{\mu}= g_v j^\mu - \frac{\xi}{6} g_v^4
({V^{\mu}})^3,
\end{equation}
\begin{equation}
(\partial_t^2-\nabla^2+m_{\rho}^2)\vec b^{\mu}= \frac{g_{\rho}}{2} \vec j^\mu 
+\frac{g_{\rho}}{2} (\vec b_{\nu} \times \vec B^{\nu \mu})
+
 g_{\rho} \partial_{\nu} (\vec b^{\nu} \times \vec b^{\mu}),
\end{equation}
\begin{equation}
(\partial_t^2-\nabla^2)A^{\mu}= \frac{e}{2} j^\mu_{em}
\end{equation}
and
$$
i \partial_t \psi= \left[ \boldsymbol{ \alpha}\cdot (-i \boldsymbol{\nabla} - g_v \mathbf{ V}
- \frac{g_{\rho}}{2} \tau_3 \mathbf{b} - e \frac{(1+ \tau_3)}{2} \mathbf{A})\right.
$$
\begin{equation}
\left.
+ \beta(M - g_s \phi) + g_v V^0
+ \frac{g_{\rho}}{2} \tau_3 b^0
 + e \frac{(1+ \tau_3)}{2} A^0 \right] \psi,
\end{equation}
where the scalar density $\rho_s$ and the baryonic current densities
are defined as
$$\rho_s=<\bar \psi \psi>,$$
$$ j^\mu=<\bar \psi \gamma^{\mu} \psi>,$$
$$\vec j^\mu =<\bar \psi \gamma^{\mu} \vec \tau \psi>,$$
$$j^\mu_{em}=<\bar \psi \gamma^{\mu} (\frac{1-\tau_3}{2}) \psi>$$
and $b_3^\mu\equiv(b^0,\mathbf{b})$.
In the static case there are no currents in the nucleus and the spatial
vector components $\mathbf  V$, $\mathbf b$ and $\mathbf  A$ are zero. Therefore, 
the equations of motion become:
\begin{equation}
\nabla^2 \phi = m_s^2\phi+{1\over2}\kappa \phi^2 
+{1\over3!} \lambda\phi^3 - g_s \rho_s,
\label{elphi} 
\end{equation}
\begin{equation}
\nabla^2 V_0 = m_v^2 V_0 +
{1\over3!}\xi g_v^4 V_0^3- g_v \rho_B,
\label{elV0}
\end{equation}
\begin{equation}
\nabla^2 b_0 =m_\rho^2 b_0 -\frac{g_\rho}{2} \rho_3,
\label{elb0}
\end{equation}
\begin{equation}
\nabla^2 A_0 =-e \rho_p,\label{elA0}
\end{equation}
where $\rho_B=\rho_p+\rho_n$ and $\rho_3=\rho_p-\rho_n$
are the barionic densities, and $\rho_p$ and $\rho_n$ are the proton 
and neutron densities.

\section{The Thomas--Fermi Approximation}

The present work is based on  the semi-classical Thomas-Fermi approximation.
In this approach
the energy of
the nuclear system with particles and anti-particles, described respectively
by the one-body phase-space distribution functions
$n_+({\mathbf{r}},{\mathbf{p}},t)$ and $n_-({\mathbf{r}},{\mathbf{p}},t)$, at position $\mathbf r$ time
$t$ and with  momentum $\mathbf p$
\begin{equation}
n_\pm({\mathbf{r}},{\mathbf{p}},t)=
\left(\begin{array}{c c}
n_{p\pm}({\mathbf{r}},{\mathbf{p}},t) &
  0 \\
 0 & n_{n\pm}({\mathbf{r}},{\mathbf{p}},t)
\end{array}\right),
\end{equation}
is (only the nuclear matter contribution and interaction terms)
\begin{equation}
E_N\,=\,\gamma\mbox{Tr}\int\frac{\d^3r\d^3p}{(2\pi)^3}\,(n_+({\mathbf r},{\mathbf p},t)
\epsilon_+\,
+n_-({\mathbf r},{\mathbf p},t)\epsilon_-)
\end{equation}
where
$$\epsilon_\pm=
\left(\begin{array}{cc}
\epsilon_{p\pm} &0\\ 0 &\epsilon_{n\pm}
\end{array}\right),
\quad
\epsilon_{i\pm}\,=\,\sqrt{(\mathbf{p}\mp {\boldsymbol{ {\cal V}}_{i}})^2 + (M-g_s
\phi)^2} 
\pm {\cal V}_{i0}, \quad i=p,n,$$
with
$$
{\cal V}_{p0}= g_v V_0  + \frac{g_\rho}{2} b_0 + e A_0\; ,
\quad {\cal V}_{n0}= g_v V_0  - \frac{g_\rho}{2}  b_0 \; ,
$$
$${\boldsymbol{{\cal V}}_{p}}= g_v {\mathbf V} +g_\rho {\mathbf b} + e {\mathbf A}\, ,
\quad {\boldsymbol{ {\cal V}}_{n}}= g_v {\mathbf V} -g_\rho {\mathbf b}\,, $$
are the classical effective one-body Hamiltonian for particles (+) and
anti-particles~(-) since particles and anti-particles have opposite 
baryonic charge and $\gamma=2$ refers to the spin multiplicity.
We can also work with the distribution function for 
particles at position $\mathbf r$, instant $t$ with momentum $\mathbf
p$,
$ f_+({\mathbf r},{\mathbf p},t)
=n_+({\mathbf r},{\mathbf p},t)$
and 
the distribution function for 
antiparticles at position $\mathbf r$, instant $t$ with momentum
$-\mathbf p$,
$ f_-({\mathbf r},{\mathbf p},t)=n_-({\mathbf r},{-\mathbf p},t) $
so that
$$
E_N\,=\,\gamma \mbox{Tr}\int\frac{\d^3r\d^3p}{(2\pi)^3}\,(f_+({\mathbf r},{\mathbf p},t)\,\,h_+
\,-f_-({\mathbf r},{\mathbf p},t)\,\,h_-)
$$
where
\ban
h_\pm&=&\pm\epsilon_\pm ({\mathbf r},\pm{\mathbf p},t)\\
&=&\left(\begin{array}{cc}
\pm\sqrt{({\mathbf p}- {\boldsymbol{{\cal V}}_{p}})^2 + (M-g_s\phi)^2} 
+ {\cal V}_{p0} & 0\\
 0 &
\pm \sqrt{({\mathbf p}- {\boldsymbol{{\cal V}}_{n}})^2 + (M-g_s\phi)^2} 
+ {\cal V}_{n0}
\end{array}\right). 
\ean
The classical entropy of a Fermi gas is given by 
\begin{equation}
S=-\gamma\sum_{i=p,n}\int\frac{\d^3rd^3p}{(2\pi)^3}\,\left(f_{i+} 
\ln\left(\frac{f_{i+}}{1-f_{i+}}\right)+
\ln(1-f_{i+}) +(f_{i+}\leftrightarrow f_{i-})\right)\;,
\end{equation}
and the thermodynamic potencial is defined as

\begin{equation}
\Omega\,=\,E-TS-\sum_{i=p,n}\mu_i B_i,\; 
\label{Omega}\end{equation}
where $B_p\,, \, B_n$ are, respectively, the proton and the neutron number:
\begin{equation}
B_i\,=\,\int\d^3r \rho_i({\mathbf r},t),\qquad \rho_i=\gamma
\int\frac{\d^3p}{(2\pi)^3}(f_{i+}-f_{i-}), \quad i=p,n\; ,
\label{rhoi}\end{equation}
$\mu_i$ is the chemical potential for particles of type $i$ and $T$ is the 
temperature.
For a system in equilibrium, the distribution functions should be chosen to
make the thermodynamic potencial $\Omega$ stationary and hence
\begin{equation}
f_{i\pm}({\mathbf r},{\mathbf p},t)\,=\,\frac{1}{1+\exp[(\epsilon\mp\nu_i)/T]}\;
,\end{equation}
where  $\nu_i\;=\mu_i-{\cal V}_{i0}$ is the effective chemical potential, 
$\epsilon=\sqrt{p^2+M^*}$ and 
$M^* =M-g_s\phi$ is the effective nucleon mass. 
In the static approximation ${\boldsymbol{{\cal V}}_\pm}=0$.

From the above expressions we get for (\ref{Omega})
$$
\Omega= \,\gamma\mbox{Tr}\int\frac{\d^3r\d^3p}{(2\pi)^3}\,(f_+({\mathbf r},{\mathbf p},t)h_
+\,-f_-({\mathbf r},{\mathbf p},t)h_-)
$$
$$
+\frac{1}{2} \int\d^3r \left [
(\nabla \phi)^2 -(\nabla V_0)^2 - (\nabla b_0)^2 - (\nabla A_0)^2 \right] 
$$
$$
+\frac{1}{2} \int\d^3r \left[ 
m_s^2 \phi^2 + \frac{2}{3!} \kappa \phi^3 + \frac{2}{4!} \lambda \phi^4
-m_v^2 V_0^2 - \frac{2}{4!} \xi g_v^4 V_0^4 
-m_\rho^2 b_0^2 \right]
$$
\begin{equation}
- \gamma T\sum_i \int\d^3r \frac{\d^3p}{(2\pi)^3} \left[ 
\frac {h_{i+}}{T} f_{i+} - \frac {h_{i-}}{T} f_{i-} + 
\ln (1 + e^{-(\epsilon - \nu_i)/T})
+ \ln (1 + e^{-(\epsilon + \nu_i)/T}) \right].
\label{pot}
\end{equation}
Equation (\ref{pot}) can then be written in the form
$$
\Omega=  \int\d^3r \left(\frac{1}{2}\left [
(\nabla \phi)^2 -(\nabla V_0)^2 - (\nabla b_0)^2 - (\nabla A_0)^2
\right] - V_{ef}\right)
$$
with
$$
V_{ef}=
-\frac{1}{2} \left[ 
m_s^2 \phi^2 + \frac{2}{3!} \kappa \phi^3 + \frac{2}{4!} \lambda \phi^4
-m_v^2 V_0^2 - \frac{2}{4!} \xi g_v^4 V_0^4 
-m_\rho^2 b_0^2  \right]$$
\begin{equation}
+ \gamma T \sum_i \int \frac{\d^3p}{(2\pi)^3} \left[  
\ln (1 + e^{-(\epsilon - \nu_i)/T})
+ \ln (1 + e^{-(\epsilon + \nu_i)/T}) \right] \label{vef}.
\end{equation}

The fields that minimize $\Omega$ satisfy the equations
\begin{equation}\frac{\partial V_{ef}}{\partial\phi}=-m_s^2\phi
-{1\over2}\kappa \phi^2 -{1\over3!} \lambda\phi^3 + g_s \rho_s,
\label{phi} \end{equation}
\begin{equation}\frac{\partial V_{ef}}{\partial V_0} =m_v^2 V_0 +
{1\over3!}\xi g_v^4 V_0^3- g_v \rho_B, \label{V0}\end{equation}
\begin{equation}\frac{\partial V_{ef}}{\partial b_0} =m_\rho^2 b_0
-\frac{g_\rho}{2} \rho_3, \label{b0}\end{equation}
\begin{equation}\frac{\partial V_{ef}}{\partial A_0}=-e \rho_p,\label{A0}\end{equation}
where $\rho_B,\, \rho_3$ were defined at the end of section 2,
$\rho_p$ has been defined in (\ref{rhoi}) and 
$$\rho_s= \gamma \sum_{i=p,n}
\int \frac{\d^3p}{(2\pi)^3}{M^*\over \epsilon}\left(f_{i+}+f_{i-}\right).$$

Comparing eqs. (\ref{elphi}--\ref{elA0})
with eqs. (\ref{phi}--\ref{A0}), we see that

\begin{equation}\nabla^2 \phi = \frac{\d^2 \phi}{\d r^2} + \frac{2}{r}\frac{\d \phi}{\d r}
=-\frac{\partial V_{ef}}{\partial\phi},
\label{phi2} \end{equation}
\begin{equation}\nabla^2 V_0 =\frac{\d^2 V_0}{\d r^2} + \frac{2}{r}\frac{\d V_0}{\d r}
= \frac{\partial V_{ef}}{\partial V_0},
\label{V02}\end{equation}
\begin{equation}\nabla^2 b_0 =\frac{\d^2 b_0}{\d r^2} + \frac{2}{r}\frac{\d b_0}{\d r}
=\frac{\partial V_{ef}}{\partial b_0},
\label{b02}\end{equation}
\begin{equation}\nabla^2 A_0 =\frac{\d^2 A_0}{\d r^2} + \frac{2}{r}\frac{\d  A_0}{\d r}
=\frac{\partial V_{ef}}{\partial A_0}.
\label{A02}\end{equation}

These coupled differential equations are solved  numerically 
and all relevant quantities (e.g. effective mass, densities, pressure),
which depend on the fields are calculated. 

\section{Two--phase coexistence}

In order to obtain the initial conditions for the program which
integrates the differential equations (\ref{phi2}-\ref{A02}) 
we determine the
conditions under which two distinct phases can coexist in infinite matter. 
In the mean field approximation 
the meson fields are replaced by their expectation values
\cite{ms,st},
\begin{equation}\phi\equiv<\phi> = \phi_0 \label{phi0},\end{equation}
\begin{equation}V^{0}\equiv<V^0> = V_0 \label{V00},\end{equation}
\begin{equation}b^{0}\equiv<b^{0}> = b_0 \label{b00},\end{equation}
and the eletromagnetic field vanishes.
The substitution of the above expressions in eqs. (\ref{elphi}),
(\ref{elV0}) and (\ref{elb0}) yields
\begin{equation}
\phi_0=- \frac{\kappa}{2 m_s^2} \phi_0^2 
-\frac{\lambda}{6}\phi_0^3 + \frac{g_s}{m_s^2} \rho_s,
\end{equation}
\begin{equation}
V_0 = - \frac{\xi g_v^4}{6 m_v^2} V_0^3 + \frac{g_v}{m_v^2} 
\rho_B,
\end{equation}
\begin{equation}
b_0 = \frac{g_\rho}{2 m_{\rho}^2} \rho_3.
\end{equation}

The thermodynamic quantities of interest are given in terms of the
above meson fields. They are the energy density:
$$ 
{\cal E}=\frac{\gamma}{2 \pi^2} \sum_{i=p,n}
\int p^2 dp \sqrt{p^2+{M^*}^2} \left(f_{i+}+f_{i-}\right)
$$
\begin{equation}
+\frac{m_v^2}{2} V_0^2 + \frac{\xi g_v^4}{8} V_0^4  
+\frac{g_\rho^2}{8 m_{\rho}^2} \rho_3^2
+\frac{m_s^2}{2} \phi_0^2 + \frac{\kappa}{6} \phi_0^3
+\frac{\lambda}{24}\phi_0^4, 
\end{equation}
the pressure:
$$ 
{ P}=\frac{\gamma}{6 \pi^2} \sum_{i=p,n}
\int \frac{p^4 dp}{\sqrt{p^2+{M^*}^2}} \left(f_{i+}+f_{i-}\right)$$
\begin{equation}
+\frac{m_v^2}{2} V_0^2 + \frac{\xi g_v^4}{24} V_0^4  
+\frac{g_\rho^2}{8 m_{\rho}^2} \rho_3^2
-\frac{m_s^2}{2} \phi_0^2 - \frac{\kappa}{6} \phi_0^3
-\frac{\lambda}{24}\phi_0^4, 
\end{equation}
the entropy density: 
\begin{equation}
{\cal S}= {1\over T}({\cal E}+{ P}-\mu_p \rho_p - \mu_n \rho_n),
 \end{equation}
and the proton fraction:
\begin{equation}
Y_p = \frac{\rho_p}{\rho_B}. 
\end{equation}

A thorough study of the possibility of phase transitions
in hot, asymmetric nuclear matter is done in \cite{barranco}
and \cite{ms}.
In figure 1 we plot the pressure in terms of the baryonic density
$\rho_B$ for each proton fraction and  for $T=10$ MeV obtained with
the MS constants. This figure is slightly
different from figure 4 of \cite{ms}, which is reproduced
when $m_s=500$ MeV. Similar behaviours are found for $T=5$ MeV and
also with the NL1 constants.

We have made use of the geometrical construction
\cite{barranco} 
in order to obtain the chemical potentials 
in the two coexisting phases for each pressure of interest.
As an example, we show $\mu_p$ and $\mu_n$ in function of the
proton fractions in figure 2 again for $T=10$ MeV and MS constants. 

In a binary system
\begin{equation}
\left( \frac{\partial \mu_p}{\partial Y_p} \right)_{T,{ P}} \ge 0
~~~{\rm and} ~~~
\left( \frac{\partial \mu_n}{\partial Y_p} \right)_{T, { P}} \le 0,
\end{equation}
known as diffusive stability, which reflects the fact that in a
stable system, energy is required to change the proton concentration
while pressure and temperature are kept constant.
In order to obtain the binodal section which contains points under
the same pressure for different proton fractions, we have used the
conditions above and
simultaneously solved the following equations:
\begin{equation}
{ P}={ P}(\nu_p,\nu_n,M^*),
\end{equation}
\begin{equation}
{ P}={ P}(\nu_p',\nu_n',{M^*}^{\prime}),
\end{equation}
\begin{equation}
\mu_i(\nu_p,\nu_n,M^*)=\mu_i(\nu_p^{\prime},\nu_n^{\prime},
{M^*}^{\prime})~~,i=p,n
\end{equation}
\begin{equation}
\frac{m_s^2}{g_s^2} \phi_0 + \frac{\kappa}{2 g_s^3} \phi_0^2 
+\frac{\lambda}{6 g_s^4}\phi_0^3 = \rho_s(\nu_p,\nu_n,M^*)
\end{equation}
and
\begin{equation}
\frac{m_s^2}{g_s^2} {\phi_0}^{\prime} + \frac{\kappa}{2 g_s^3} {\phi_0^2}
^{\prime} 
+\frac{\lambda}{6 g_s^4}{\phi_0^3}^{\prime} = \rho_s(\nu_p^{\prime},
\nu_n^{\prime},{M^*}^{\prime}).
\end{equation}

The binodal sections
for the  MS and the NL1  constants and temperatures equal to 5 and 10 MeV
 are plotted, respectively, in figure 3 and
in figure 4.
For certain values of proton and neutron chemical
potentials, the system may be at the same pressure with different
densities  and proton concentrations,
 which allows for the possibility of phase transitions.
For the sake of completeness, we also show in tables 2, 4 and 6
some of the points taken from the  binodal sections.
The results we have chosen as input to the code which solves
the differential equations (\ref{phi2}-\ref{A02})
are displayed in the last three columns of
these tables.

\section{Numerical Results}

Solving numerically the set of coupled equations 
(\ref{phi2}--\ref{A02})
is not trivial. 
The main problem is related with the boundary conditions which have
to be set within the droplet. 
To understand better
this statement, please refer to the Appendix, where the formulae are simplified 
to the $T=0$ case and this problem becomes clear. At first, we have tried 
to use the code COLSYS \cite{colsys}, as suggested in \cite{sed} and \cite{md}, 
but we have obtained
satisfactory results only for symmetric nuclear matter \cite{vinas}.
We have finally opted for another code, written with the help of the
Gears stiff integration method, which uses as 
input the
temperature, the size of the mesh, boundary conditions and initial
conditions. The chemical potentials are output and can be
{\it fitted} in accordance with the size of the mesh. In the Appendix
the boundary and initial conditions we have used are explicitly 
written.

The radius $R_{max}$ (see Appendix) fixes the neutron chemical potential
calculated at the last mesh point. The proton chemical potential
is fixed by   another (inner) mesh point, which  depends on 
the difference desired between both potentials. If $\mu_p=\mu_n$
(symmetric nuclear matter), $R_{max}$ fixes both chemical potentials at the
last mesh point. We have considered that convergence has been achieved
when the baryonic density does not vary more than 0.5 per cent and the
chemical potentials are close to the ones obtained from the binodal
section for a certain proton fraction.

As an example, in figure 5 we plot the fields which are solutions of the coupled equations 
for the MS parametrization at T=10 MeV
with the initial conditions given in table 2 for $Y_p=0.3$.
The corresponding barionic density is plotted in figure 6 and  represents a droplet
of the liquid phase (small $r$) in the background of the vapour phase
(large $r$). 
Similar density profiles are obtained for all other proton fractions,
except for $Y_p=0.5$, when the curves for proton and neutron densities
coincide. 
The  region at the surface with extra neutrons is known as 
{\it neutron skin}. Some quantities of interest to study the surface properties 
are the  two {\it squared--off} radii $R_n$ and
$R_p$, defined in \cite{rpl} as 
\begin{equation}
\int_0^{R_{max}} \rho_n(r) \d r = \rho_{n,i} R_n+ \rho_{n,f}(R_{max}-R_n),
\label{raion}
\end{equation}
and 
\begin{equation}
\int_0^{R_{max}} \rho_p(r) \d r = \rho_{p,i}R_p + \rho_{p,f}(R_{max}-R_p),
\label{raiop}
\end{equation}
where $\rho_i$ refers to the liquid density, $\rho_f$ to the gas 
density, $R_{max}$ is the size of the mesh for which convergence is
achieved,
and the neutron skin thickness \cite{cev}
\begin{equation}
\Theta=R_n-R_p.
\end{equation}
 These quantities are computed  for the droplet solutions we
obtain and given in tables 3, 5, 7.

The droplet surface energy and thickness are
obtained from the free energy of a system with a fixed number of
particles $B=B_p+B_n$, in which a droplet of arbitrary size grows
in the background of the vapour phase. 
Within the small surface thickness approximation it reads \cite{np}:
\begin{equation}
F=\int 4\pi r^2 \d r 
\left[ \left(\frac{\d \phi}{\d r}\right)^2- 
\left(\frac{\d V_0}{\d r}\right)^2 -
\left(\frac{\d b_0}{\d r}\right)^2-
\left(\frac{\d A_0}{\d r}\right)^2 -C \right]+\mu_p B_p +\mu_n B_n.
\end{equation}
For droplets with radius $R$ and small surface thickness,
\begin{equation}
F(R)=4 \pi \sigma R^2 - CV + \mu_p B_p +\mu_n B_n, \label{free}
\end{equation}
where $C$ is a constant and $V$ is the volume of the system. 
The surface energy per unit area of these droplets is
\begin{equation}
\sigma=\int_0^\infty \d r \left[ \left(\frac{\d \phi}{\d r}\right)^2- 
\left(\frac{\d V_0}{\d r}\right)^2 -
\left(\frac{\d b_0}{\d r}\right)^2-
\left(\frac{\d A_0}{\d r}\right)^2 \right].
\label{sig}\end{equation}
The surface thickness $t$ is defined as the width of the region where
the density drops from $0.9 \rho_{B0}$ to $0.1 \rho_{B0}$, where $\rho_{B0}$ 
is the baryonic density at $r=0$.  
According to \cite{bb}, for $T=0$,
$\sigma$ should be of the order of 1.25 
MeV fm$^{-2}$ and $t$ of the order of 2.2 fm.

In table 2 some points taken from the binodal section  for the MS constants
at $T= 10$ MeV are explicitly written. 
In table 3
results found for the proton fractions at $r=0$ ($Y_p(i)$),
chemical potentials, the surface energy, its thickness, the size of
the mesh for which convergence is achieved, $R_{max}$,  $R_p$ and the neutron skin thickness 
are displayed. 
Notice that there is a small
discrepancy between the  proton and neutron chemical potentials
given in table 2 and the ones displayed in table 3. This is due to 
finite size effects and the inclusion of the Coulomb interaction.
In tables 4 and 5 again the points obtained from the binodal
section at $T=5$ MeV and the respective droplet solutions are shown
for the MS set of constants while in tables 6 and 7 the NL1 constants
are used. 

At this point some comments are in order. 
The size of the mesh ($R_{max}$) given in tables 3, 5 and 7 are the smaller 
values for which there is convergence for a given density at $r=0$ and 
a given proton fraction $Y_p$ within the accuracy of the present
numerical calculations, i.e., $\pm 0.5\%$. We have chosen to compare data
corresponding to the same value of the
  proton fraction at $r=0$  because this parameter is
independent of the propeties of the surface.
A larger mesh size would converge to a larger droplet with the same
values for the
proton fraction $Y_p$ and the density at $r=0$, and the same chemical potentials.
For a detailed explanation on the introduction of the
Coulomb field, please refer to the Appendix.
Before drawing our conclusions, we would like to emphasize that in our 
calculations, the proton and neutron numbers are never fixed. They are 
just consequence of the results for the fields and densities obtained from
the convergent solutions of the differential equations. 

\section{Conclusions}

We first examine the behaviour of the total baryonic density.
From figure 6, one can see that it falls from the initial liquid density
to the vapour density, which is very small, but different from zero,
as expected. One can compare this figure with the densities
presented in table 2 for infinite nuclear matter with $Y_p=0.3$.
Concerning the neutron and proton densities, figure 6 shows the
same profile obtained in \cite{suraud} and \cite{rpl} with
non--relativistic models. 

The proton fraction in the vapour phase 
is smaller than in the liquid phase ($Y_p(i)$),
except for the symmetric nuclear matter, when it remains
unaltered. This can be seen from tables 2, 4 and 6
and confirmed for the droplet solutions. This fact 
can be interpreted as a nucleus with a given proton concentration
(the phase of higher density) in equilibrium with a gas of
drip nucleons, mostly neutrons (the lower density phase) with
a much smaller proton concentration.

From tables 3, 5 and 7, one can check that
the surface energy $\sigma$ increases with the initial proton
fraction and its thickness $t$ decreases. In fact, the larger the  proton
fraction the less important is the contribution from the $b_0$ field 
in the $\sigma$ calculation as can be seen from (\ref{sig}).
For symmetric nuclear
matter, the results are compatible with the ones suggested
in \cite{bb} and for asymmetric matter the surface thickness
results are  comparable with the ones presented in 
\cite{cev}.
One can also see that the larger the proton fraction, the
smaller the size of the mesh for which convergence is
achieved. This may be due to the decrease of neutron--proton asymmetry
and therefore, the increase of the droplet binding.
Some conclusions with respect to the temperature dependence of the
droplet solutions can be drawn  comparing tables 3 and 5. First, it is easy to 
note that the size of the mesh must  be larger for
higher temperatures. A similar statement was made in \cite{md}.
Also, the surface energy decreases with the increase of temperature, 
while the surface thickness is larger for  higher temperatures. 
The decrease of $\sigma$ with the 
temperature is easily understood from (\ref{sig}), because for higher
temperatures the fields decrease more smoothly
 and spread out over a larger distance 
at the surface and, therefore, their derivatives are smaller. 
  
In the same tables the squared-off proton radius and the 
 neutron skin thickness are also shown.
Both quantities decrease with the increase of the  proton
fraction at $r=0$. The numbers we have obtained can be compared with the
ones found in \cite{rpl}. These behaviours could be a consequence
of the increase of the droplet binding. The neutron skin thickness is a quantity
which is larger for lower temperatures, except for symmetric nuclear
matter, when it is zero,  independently of the temperature considered.

Concerning the importance of the Coulomb interaction and its
consequences in the droplet formation, one can see, from figure
6, that the proton and neutron densities are indeed modified
by the eletromagnetic field, as pointed out  in
\cite{blv}. Nevertheless, for the same value of $Y_p$ at $r=0$,
the profile of the surface with eletromagnetic field is almost
unaltered with respect to the results obtained without the Coulomb interaction.
This fact is reflected in the
results displayed in table 3.a, which were obtained with the
same input parameters as the ones shown in the third line of table 3,
but without the inclusion of the eletromagnetic interaction.
The surface energy $\sigma$, the proton radius $R_p$ and the neutron
skin $\Theta$ are slightly 
different when they are calculated with and without the Coulomb 
potential while   $t$ is practically the same.
 Two effects of the eletromagnetic field in the present
calculation is to decrease the number of particles in the
droplet, since the central density becomes smaller, and to
increase the proton radius $R_p$.

The conclusions drawn above are independent of the sets of parameters
used.

To the best of our knowledge, this is the first time
that the eletromagnetic interaction is taken into account within
the framework of a relativistic model in order to calculate
surface properties. We believe that a more systematic study of
the importance of the Coulomb field for various temperatures and
proton fractions has still to be made and this problem will be
tackled in a forthcoming work. 

\section*{Appendix}

For $T=0$ the distribution functions are simply given by:
$$f_i=\theta(k_{Fi}^2(r)-p^2), ~~~~~~~i=p,n$$
and, hence,
$$h_i=h_{i+},$$
\ban
V_{ef}&=&
-\frac{1}{2} \left[ 
m_s^2 \phi^2 + \frac{2}{3!} \kappa \phi^3 + \frac{2}{4!} \lambda \phi^4
-m_v^2 V_0^2 - \frac{2}{4!} \xi g_v^4 V_0^4 
-m_\rho^2 b_0^2  \right]\\
& &- \gamma  \sum_{i=p,n} \int \frac{\d^3p}{(2\pi)^3} h_i f_i 
+ \mu_p \rho_p+\mu_n \rho_n,
\ean
$$
\rho_s= \gamma \sum_{i=p,n}
\int \frac{\d^3p}{(2\pi)^3}{M^*\over \epsilon}f_{i} =
\frac{\gamma}{2\pi^2} \sum_{i=p,n} \int_0^{k_{Fi}(r)} 
p^2 \d p {M^*\over \epsilon},
$$
$$
\rho_i= \gamma 
\int \frac{\d^3p}{(2\pi)^3}f_{i}=\frac{\gamma}{6\pi^2}k_{Fi}^3(r).
$$

Minimization of $\Omega$ with respect to $ k_{Fi}(r),\, i=p,n$,  
gives rise to the following conditions
$$
k_{Fp}^2(r)\left(\mu_p-\sqrt{k_{Fp}^2(r)+{M^*}^2(r)}-
g_v V_0- \frac{g_\rho}{2} b_0 -e A_0\right)=0
$$
and
$$
k_{Fn}^2(r)\left(\mu_n-\sqrt{k_{Fn}^2(r)+{M^*}^2(r)}-
g_v V_0+ \frac{g_\rho}{2} b_0\right)=0.
$$
We obtain $k_{Fp}(r)=0$ and $k_{Fn}(r)=0$ or,
for $k_{Fp}(r)$ and $k_{Fn}(r)$  different from zero,
\begin{equation}
\mu_p=\sqrt{k_{Fp}^2(r)+{M^*}^2(r)}+ g_v V_0  + g_\rho  b_0 + 
e A_0,\label{mup}\end{equation}
\begin{equation}
\mu_n=\sqrt{k_{Fn}^2(r)+{M^*}^2(r)}+ g_v V_0  - g_\rho  b_0.
\label{mun}\end{equation}
The value of $k_{Fp}(r)$ and $k_{Fn}(r)$ is obtained inverting these
last two equations.
The  discontinuity on the values of $k_{Fp}(r)$ and $k_{Fn}(r)$ has to be 
taken into account in the code
that solves the differential equations (\ref{phi}--\ref{A0}). 
Outside the droplet $k_{Fp}(r)$ and $k_{Fn}(r)$ are zero and   the mesons are free. 

In our code, the boundary conditions are given by
$$\frac{\d \phi}{\d r}(r=0)=\frac{\d V_0}{\d r}(r=0)=
\frac{\d b_0}{\d r}(r=0)=\frac{\d A_0}{\d r}(r=0)= 0$$
and for $r=R_{max}$, where $R_{max}$ is the size of the mesh,
$$\frac{\d \phi}{\d r}+(m_s+\frac{1}{R_{max}})\phi=0,$$
with similar equations for $V_0$ and $b_0$ and
$$\frac{\d A_0}{\d r}+\frac{A_0}{R_{max}}=0,$$
or, considering the electron screening, 
$A_0$ is zero at the last point of the mesh,
$$A_0(R_{max})=0.$$
Both boundary conditions give similar results.

As initial guesses for the meson fields we have used Fermi like functions such as
$$\phi(r)=\frac{\phi_0}{1+\exp(m_s(r-0.8R_{max}))},$$ 
for the scalar field.  The values for $r=0$, i.e.  $\phi_0$, $V_{00}$ and $b_0$
were obtained from the
binodal section.
An initial guess for the eletromagnetic field is the field of a homogeneous 
spherical
distribution of protons which at $r=0$ is of the order of $1 \times 10^{-2}$ in
units of nucleon mass. The convergence is obtained for different
initial guesses except if the initial guesses are much stronger than
the value refered above.
 We also
suppose that the droplets are formed in an electrically neutral enviroment, 
as we would find in
neutron stars.

\vskip 0.35in
\begin{center}
{\bf Acknowledgements}
\end{center}
We would like to thank Dr. Manuel Fiolhais for helping us with
the COLSYS code, Dr. Hans Walliser for giving us his fortran
code which solves the differential equations
and for his useful suggestions and Dr. J. da Provid\^encia
for his comments. One of the authors (DPM) would like to acknowledge
the warm hospitality of the Centro de F\'{\i}sica Te\'orica of the
University of Coimbra, where this work was acomplished.
This work has been partially supported by Capes - Brazil.

\newpage
\section{ Figure Captions}

\begin{itemize}
\item{\bf Figure 1.} 
The pressure in terms of the baryonic density is plotted
for each proton fraction for the MS set of constants. 
The first curve on the left relates to
$Y_p=0$, the second one to $Y_p=0.1$ ,..., the last one to $Y_p=0.5$.
The temperature is $T=10$ MeV. 
The pressure is given in
MeV/fm$^3$ and the density in fm$^{-3}$.

\item{\bf Figure 2.} The proton (lower curve) and the neutron (upper curve)
chemical potentials are shown in function of the proton fraction for
the pressure of 0.12 MeV/fm$^3$. Again this graph is plotted for MS constants
and $T=10$ MeV.

\item{\bf Figure 3.} Binodal section for $T=10$ (dashed line) and $T=5$ MeV
(solid line) with the MS constants. The pressure is given in
MeV/fm$^3$.

\item{\bf Figure 4.} Binodal section for $T=10$ (dashed line) and $T=5$ MeV
(solid line) with the NL1 constants. The pressure is given in
MeV/fm$^3$.

\item{\bf Figure 5.} $\phi$,$V_0$,$A_0$ and $b_0$ (in this order, 
from top to bottom)
are shown in terms of $r$. The fields are given in nucleon mass units and are
obtained with the input values of table 2 for $Y_p=0.3$. 

\item{\bf Figure 6.} From top to bottom the density profiles for
the baryons $\rho_B(r)$, the neutrons $\rho_n(r)$ and the protons
$\rho_p(r)$ in fm$^{-3}$ are plotted for the same 
case as in Fig. 5. The solid curves were obtained with the inclusion of the
eletromagnetic field and the dashed ones without it.
\end{itemize}

\section{Table Captions}

\begin{itemize}

\item{\bf Table 1.} Sets of parameters used in this work. 
All masses are given in MeV. The * is a reminder that the
authors of \cite{ms} use a different scalar meson mass.

\item{\bf Table 2.} Results obtained from the binodal section
built with the MS constants for $T=10$ MeV.
The pressures are given in MeV/fm$^3$,
the densities in fm$^{-3}$, 
the chemical potentials in MeV and the fields in units of nucleon mass.

\item{\bf Table 3.} 
Output results given by the solution of the coupled differential
equations with the MS constants and $T=10$ MeV. 
Index $i$ refers to $r=0$.
The effective chemical potentials are given in units of nucleon mass, the
chemical potentials in MeV, the surface energy in
MeV/fm$^2$ and the surface thickness in fm. $R_{max}$ is the size
of the mesh for which convergence is achieved and it is given in
fm. $R_p$ and $\Theta$ are also given in fm.

\item{\bf Table 3.a}
Output results given by the solution of the coupled differential
equations with the MS constants and $T=10$ MeV without the
inclusion of the eletromagnetic field.
The units are the same as in Table 3.

\item{\bf Table 4.} Results obtained from the binodal section
built with the MS constants for $T=5$ MeV. 
The units are the same as in Table 2.

\item{\bf Table 5.} 
Output results given by the solution of the coupled differential
equations with the MS constants and $T=5$ MeV. 
The units are the same as in Table 3.

\item{\bf Table 6.} Results obtained from the binodal section
built with the NL1 constants for $T=5$ MeV. 
The units are the same as in Table 2.

\item{\bf Table 7.} 
Output results given by the solution of the coupled differential
equations with the NL1 constants and $T=5$ MeV. 
The units are the same as in Table 3.
\end{itemize}
\newpage

\begin{landscape}

\centerline{\bf Table 1 - parameters}
\vspace{0.5cm}
\begin{tabular}{|c|c|c|c|c|c|c|c|c|c|c|c|}
\hline
Force & [ref] & $C_s^2$ & $C_v^2$ & $C_{\rho}^2$ & $\kappa/M \times 10^{-3}$ 
& $\lambda \times 10^{-3}$ & $\xi$ 
& $M$ & $m_s$ & $m_v$ & $m_{\rho}$ \\
\hline
MS  & [1] & 374.770 & 260.570 & 106.91 & $3.0809 \times 10^{3}$ 
& $8.106 \times 10^{3}$ & 0.02364 
& 939. & 550.00* & 783.00 & 770.00 \\
NL1 & [12] & 373.176 & 245.458 & 149.67 & 2 $g_s^3~~2.4578 $
& - 6 $g_s^4~~3.4334$ & 0.0 & 938. & 492.25 & 795.36 &
763.00 \\
\hline
\end{tabular}
\end{landscape}
\newpage
\renewcommand{\arraystretch}{1.0} 

\centerline{\bf Table 2 }
\vspace{0.5cm}

\begin{tabular}{|c|c|c|c|c|c|c|c|}
\hline
$Y_p$& $\cal P$  & $\mu_p$ & $\mu_n$ & $\rho_B$ & $\phi_0
\times 10^{-2}$& $V_{00}\times 10^{-2}$&$b_0\times 10^{-2}$ \\
\hline
0.10 & 0.38 &   873.90 & 941.36 & 0.08 & 2.01
& 1.30&-0.35\\
0.18& & & & 0.10 & 2.69&1.76&-0.39\\
\hline
0.20 &0.35 & 874.66& 940.86& 0.11 & 2.82& 1.85&-0.38\\
0.08& & & &  0.07 &1.82&1.17&-0.33\\
\hline
0.25& 0.27&877.55&939.04&0.13&3.15&2.07&-0.36\\
0.05&     &  &&0.05 &1.35&0.85&-0.25\\
\hline
0.30 & 0.18 & 882.16 &936.42& 0.14& 3.44& 2.28
&-0.33\\
0.03&  & & & 0.03 &0.83 &0.51&-0.16\\
\hline
0.35 &  0.10&888.56&932.55&0.15&3.69 &2.44
&-0.26\\
0.03&      & & &0.01&0.41&0.25&-0.08\\
\hline
0.40 &  0.06& 896.36& 927.37& 0.16& 3.87& 2.57
&-0.19\\
0.06&  &     && 0.01 &0.21& 0.12&-0.04\\
\hline
0.45 &   0.03&904.90&920.77&0.17&3.98 &
2.64 &-0.10\\
0.19&  & & &$4.\times 10^{-3}$ &0.11 &0.07&-0.01\\
\hline
0.50&0.03&913.20&913.18& 0.17&4.01&2.67 &-1.0$\times10^{-4}$\\
0.50& &   & &$3.\times 10^{-3}$ &0.09 & 0.05&-0.1$\times10^{-4}$\\
\hline
\end{tabular}

\vspace{0.5cm}
\centerline{\bf Table 3}
\vspace{0.5cm}
 
\begin{tabular}{|c|c|c|c|c|c|c|c|c|c|}
\hline
$Y_p(i)$ & $\nu_p(i)$ & $\nu_n(i)$ & $\mu_p$ & $\mu_n$ & $\sigma$ & $t$
& $R_{max}$ & $R_p$ & $\Theta$\\
\hline
0.15 &  0.75170 & 0.78836 & 876.18 & 935.30 & 0.22 
& 5.03 & 16.70 & 12.46 & 0.75\\

0.23 &  0.70667 & 0.73960 & 879.62 & 935.25 & 0.35
& 4.72 & 15.73 & 11.63 & 0.71\\

0.30 &  0.66064 & 0.68956 & 884.29 & 933.99 & 0.58 
& 3.36 & 10.49 & 7.50 & 0.59\\

0.36 &  0.62472 & 0.64845 & 892.46 & 932.22 & 0.70
& 3.21 & 9.44 & 6.75 & 0.47\\

0.38 &  0.61380 & 0.63467 & 896.94 & 931.08 & 0.76 
& 3.02 & 8.39 & 5.96 & 0.38\\

0.50 & 0.61306 & 0.61328 & 913.20 & 913.20 & 1.21 
& 2.46 & 7.34 & 3.72 & 0.0\\
\hline
\end{tabular}

\vspace{0.5cm}
\centerline{\bf Table 3.a}
\vspace{0.5cm}
 
\begin{tabular}{|c|c|c|c|c|c|c|c|c|c|}
\hline
$Y_p(i)$ & $\nu_p(i)$ & $\nu_n(i)$ & $\mu_p$ & $\mu_n$ & $\sigma$ & $t$
& $R_{max}$ & $R_p$ & $\Theta$\\
\hline
0.30 &  0.65717 & 0.68534 & 882.72 & 933.99 & 0.60 
& 3.36 & 10.49 & 7.39 & 0.62\\
\hline
\end{tabular}

\newpage

\centerline{\bf Table 4}
\vspace{0.5cm}

\begin{tabular}{|c|c|c|c|c|c|c|c|}
\hline
$Y_p$& $\cal P$  & $\mu_p$ & $\mu_n$ & $\rho_B$ & $\phi_0
\times 10^{-2}$& $V_{00}\times 10^{-2}$& $b_0\times 10^{-2}$ \\
\hline
0.15 & 0.52 & 868.12 & 948.93 & 0.12 & 
3.01 & 1.97 & -0.48\\
0.05 &       &  & & 0.09 &
2.28 & 1.48 & -0.45\\
\hline
0.20 & 0.37  & 871.18 & 946.76 & 0.13 &
3.26 & 2.15 & -0.46\\
0.02 &       &  & & 0.07 &
1.92 & 1.23 & -0.39\\
\hline
0.25 & 0.19  & 875.63 & 943.75 & 0.14&
3.49 & 2.31 & -0.41\\
$6. \times 10^{-3}$ &      &  & & 0.05 &
1.46 & 0.93 & -0.30\\
 \hline
0.30 & 0.06 & 881.47 & 940.25 & 0.15 &
3.73 & 2.47 & -0.36\\
$2. \times 10^{-4}$ &       & & & 0.02 &
0.62 & 0.38 & -0.12\\
\hline
0.35 & 0.02  & 888.80 & 936.36 & 0.17 &
3.96 & 2.63 & -0.29\\
$2. \times 10^{-4}$&      &  & & $4. \times 10^{-3}$  &
0.13 & 0.08 & -0.03\\
\hline
0.40 & 6.  & 897.55 & 930.92 & 0.18 &
4.12 & 2.74 & -0.20\\
$2. \times 10^{-3}$ &$\times 10^{-3}$ & & & $1. \times 10^{-3}$ &
0.04 & 0.02& -0.01\\
\hline
0.45 & 1.  & 906.87 & 924.01 & 0.18 &
4.22 & 2.81 & -0.10\\
0.03 & $\times 10^{-3}$ & & & $3. \times 10^{-4}$ & 
0.01 & 0.01 & -$2. \times 10^{-3}$\\
\hline
0.50  & 5. & 915.82 & 915.89 & 
0.18 & 4.25 & 2.83 & 0.0\\
0.50 & $\times 10^{-4}$ & & & $1. \times 10^{-4}$  &
$3. \times 10^{-3}$  &$2. \times 10^{-3}$  & 0.0 \\
\hline
\end{tabular}

\vspace{0.5cm}
\centerline{\bf Table 5}
\vspace{0.5cm}

\begin{tabular}{|c|c|c|c|c|c|c|c|c|c|}
\hline
$Y_p(i)$ & $\nu_p(i)$ & $\nu_n(i)$ & $\mu_p$ & $\mu_n$ & $\sigma$ & $t$
& $R_{max}$ & $R_p$ & $\Theta$\\
\hline
0.10 & 0.75486 & 0.79229 & 873.60 & 936.92 & 0.22 & 3.27 
& 12.59 & 8.63 & 1.16\\

0.22 & 0.70784 & 0.73992 & 879.47 & 936.02 & 0.42 & 2.94 
& 10.49 & 7.02 & 0.98\\

0.28 & 0.67424 & 0.70386 & 882.57 & 935.11 & 0.58 & 2.77 
& 9.23 & 6.15 & 0.81\\

0.38 & 0.91239 & 0.63308 & 899.08 & 934.24 & 0.92 & 2.49 
& 7.34 & 4.74 & 0.50\\

0.50 & 0.59357 & 0.59381 & 915.80 & 915.80 & 1.35 & 2.33 
& 5.24 & 3.69 & 0.0\\
\hline 
\end{tabular}

\newpage

\centerline{\bf Table 6}
\vspace{0.5cm}

\begin{tabular}{|c|c|c|c|c|c|c|c|}
\hline
$Y_p$& $\cal P$  & $\mu_p$ & $\mu_n$ & $\rho_B$ & $\phi_0
\times 10^{-2}$& $V_{00}\times 10^{-2}$ & $b_0\times 10^{-2}$ \\
\hline
0.10 & 0.29 & 885.36 & 946.59 & 0.07 &
1.90 & 1.11 & -0.35\\
0.11 &        &  & & 0.06 & 
1.82 & 1.06 & -0.34\\
\hline
0.15 & 0.27  & 885.91 & 946.13 & 0.07 & 
2.10 & 1.24 &-0.35\\
0.07 &        & & & 0.05 &
1.61 & 0.92 & -0.32\\
\hline
0.20 & 0.24 & 886.59 & 945.65 & 0.09 &  
2.43 & 1.45 & -0.35\\
0.06 &        &  &  & 0.05 &  
1.51 &  0.86 & -0.31\\
\hline
0.25 & 0.18 & 888.97 & 944.02 & 0.10 & 
2.73 & 1.64 & -0.33\\
0.03 &    & &  & 0.04 &  
1.23 &  0.69 & -0.26\\
\hline
0.30 & 0.10 & 892.61 &  941.56 & 0.11 &  
3.02 &  1.83 & -0.30\\
0.01 &    & &  & 0.03 & 
0.85 &  0.47 & -0.19\\
\hline
0.35 &  0.04  & 897.60 & 938.35  & 0.12 & 
3.35 & 2.05 & -0.25\\
$3. \times 10^{-3}$ &  & & & 0.01 & 
0.36 & 0.19 & -0.08\\
\hline
0.40 &  0.01 &  904.32 &  933.97 &  0.13 &  
3.67  & 2.26 & -0.18\\
$5. \times 10^{-3}$ &        &  & & $3. \times 10^{-3}$  &  
0.10 &  0.05 & -0.02\\
\hline
0.45  & 4.  &  912.06 &  928.06 &  0.14 &  
3.89 & 2.40 & -0.10\\
0.05 & $\times 10^{-3}$  &  &  &$8. \times 10^{-4}$  &  
0.03 &  0.01 & -$5. \times 10^{-3}$\\
\hline
0.50  & 1.65 &  920.42 &   920.46 &  0.14 &   
3.97 &   2.45 &-$2. \times 10^{-4}$ \\
0.50   &$\times 10^{-3}$     & & & 
$3. \times 10^{-4}$ &   0.01  &  0.01 &  0.0\\
\hline
\end{tabular}

\vspace{0.5cm}
\centerline{\bf Table 7}
\vspace{0.5cm}
 
\begin{tabular}{|c|c|c|c|c|c|c|c|c|c|}
\hline
$Y_p(i)$ & $\nu_p(i)$ & $\nu_n(i)$ & $\mu_p$ & $\mu_n$ & $\sigma$ & $t$
& $R_{max}$ & $R_p$ & $\Theta$\\
\hline
0.15 &  0.81150 & 0.83765 & 885.55 & 937.21 & 0.21 
& 3.55 & 13.64 & 9.56 & 1.02\\
0.24 &  0.77241 & 0.79539 & 889.32 & 936.84 & 0.34
& 3.27 & 12.59 & 8.84 & 0.86\\
0.32 &  0.72522 & 0.74488 & 895.33 & 936.27 & 0.56
& 3.00 & 11.54 & 8.26 & 0.64\\
0.38 &  0.68604 & 0.70168 & 902.60 & 934.54 & 0.81
& 2.65 & 9.44 & 6.80 & 0.44\\
0.42 &  0.65833 & 0.67021 & 909.25 & 932.60 & 1.00
& 2.51 & 8.39 & 6.15 & 0.28\\
0.45 & 0.63661 & 0.64454 & 914.91 & 929.43 & 1.17 
& 2.50 & 7.34 & 5.45 & 0.15\\

\hline
\end{tabular}

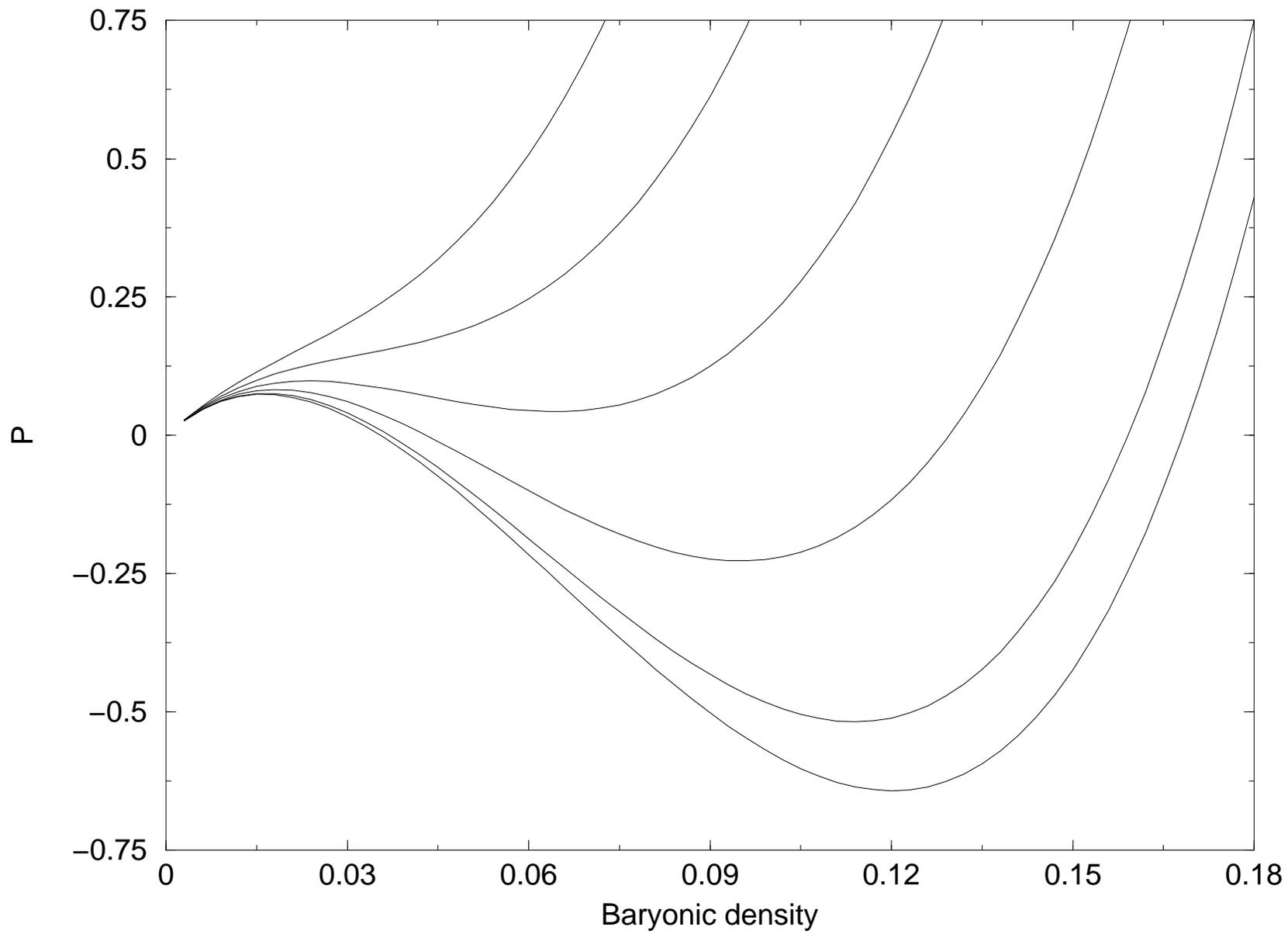
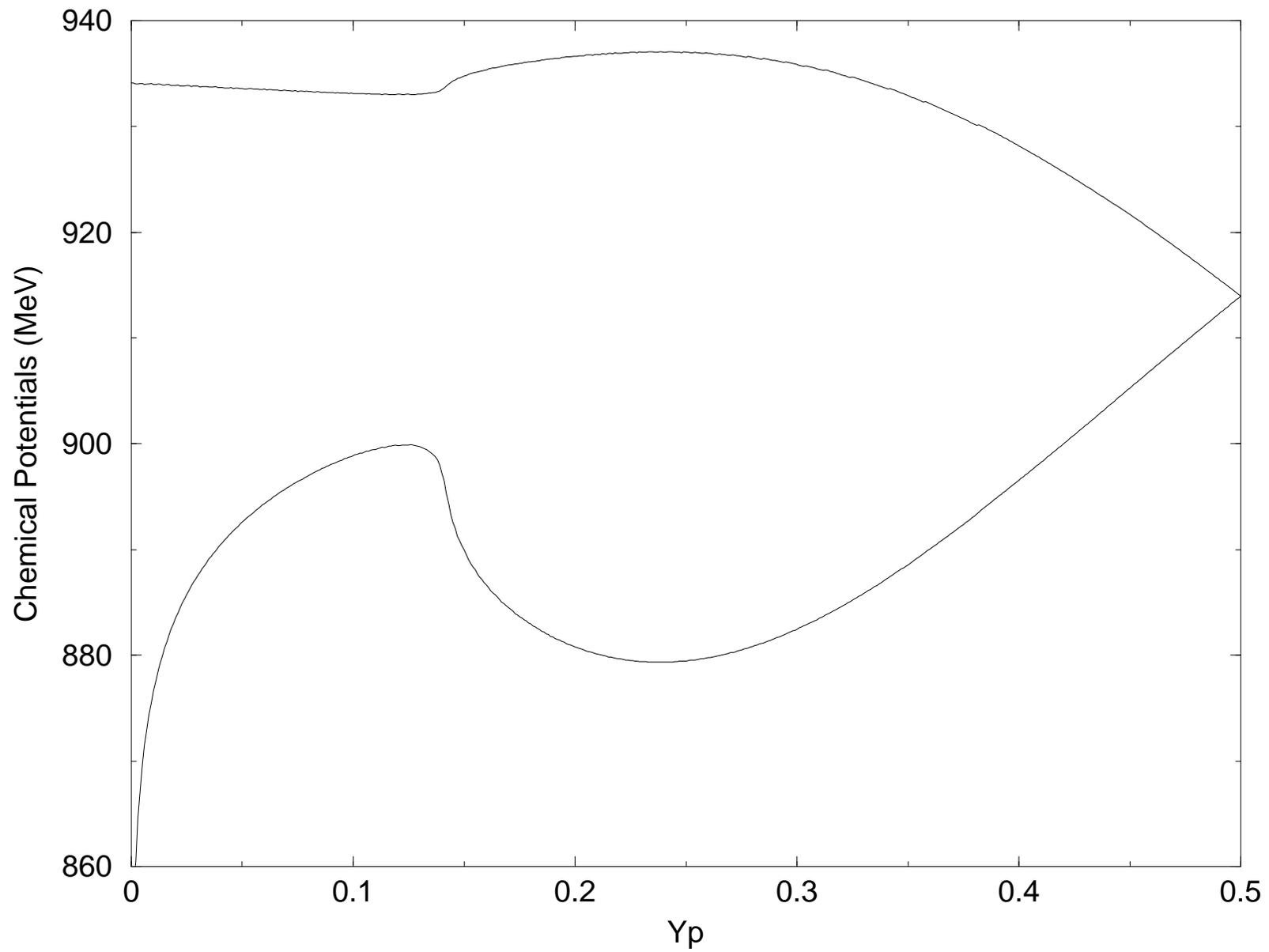
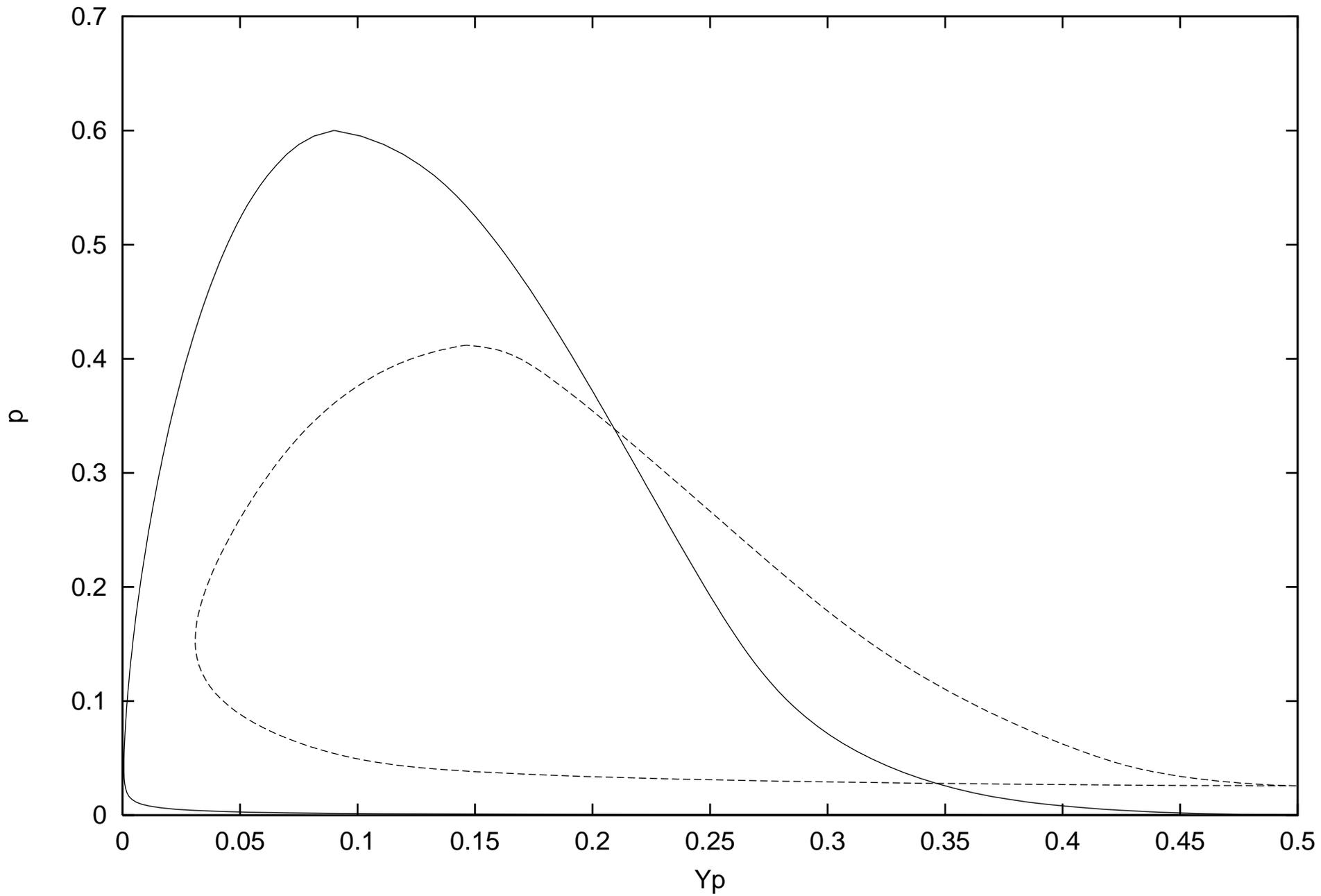
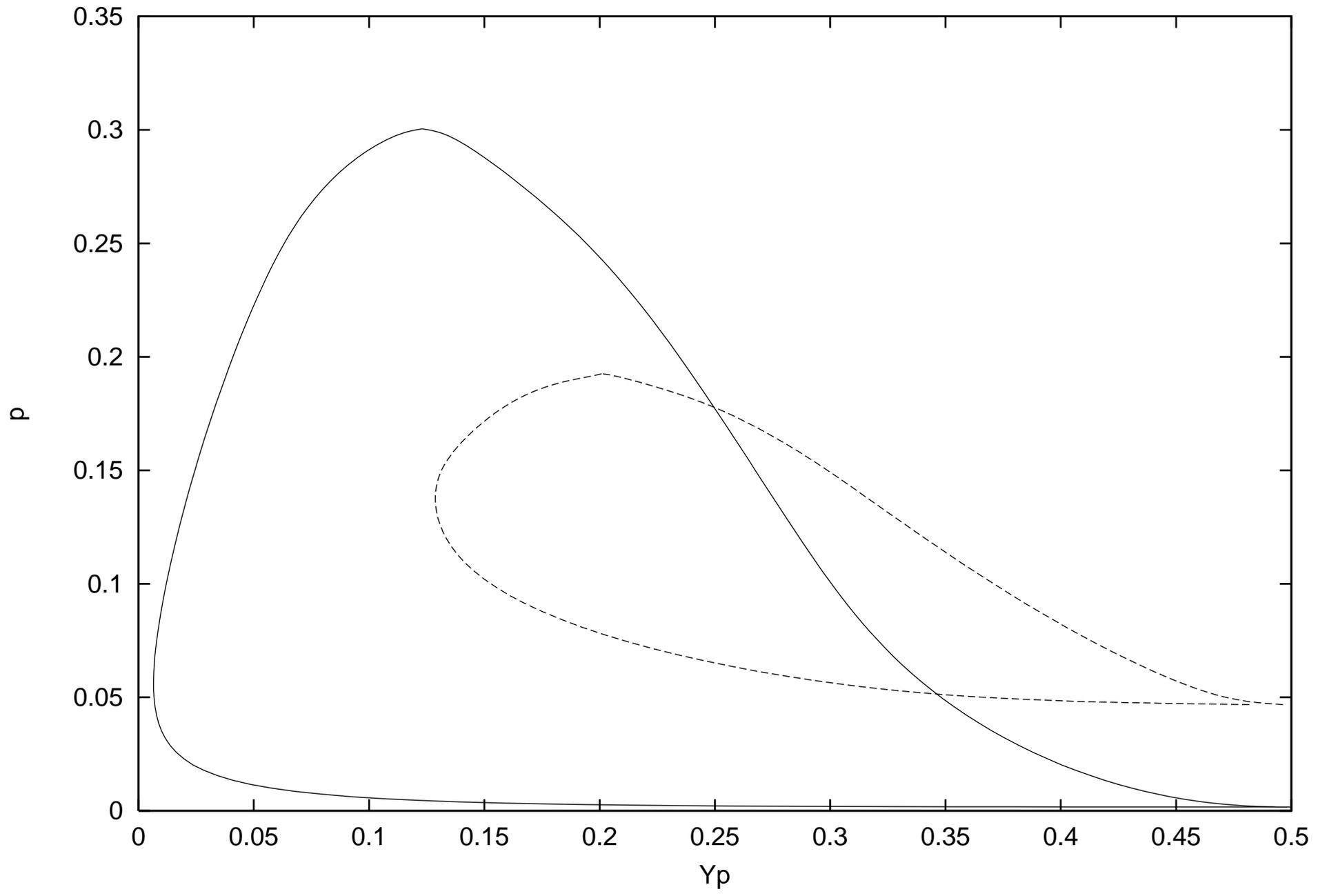
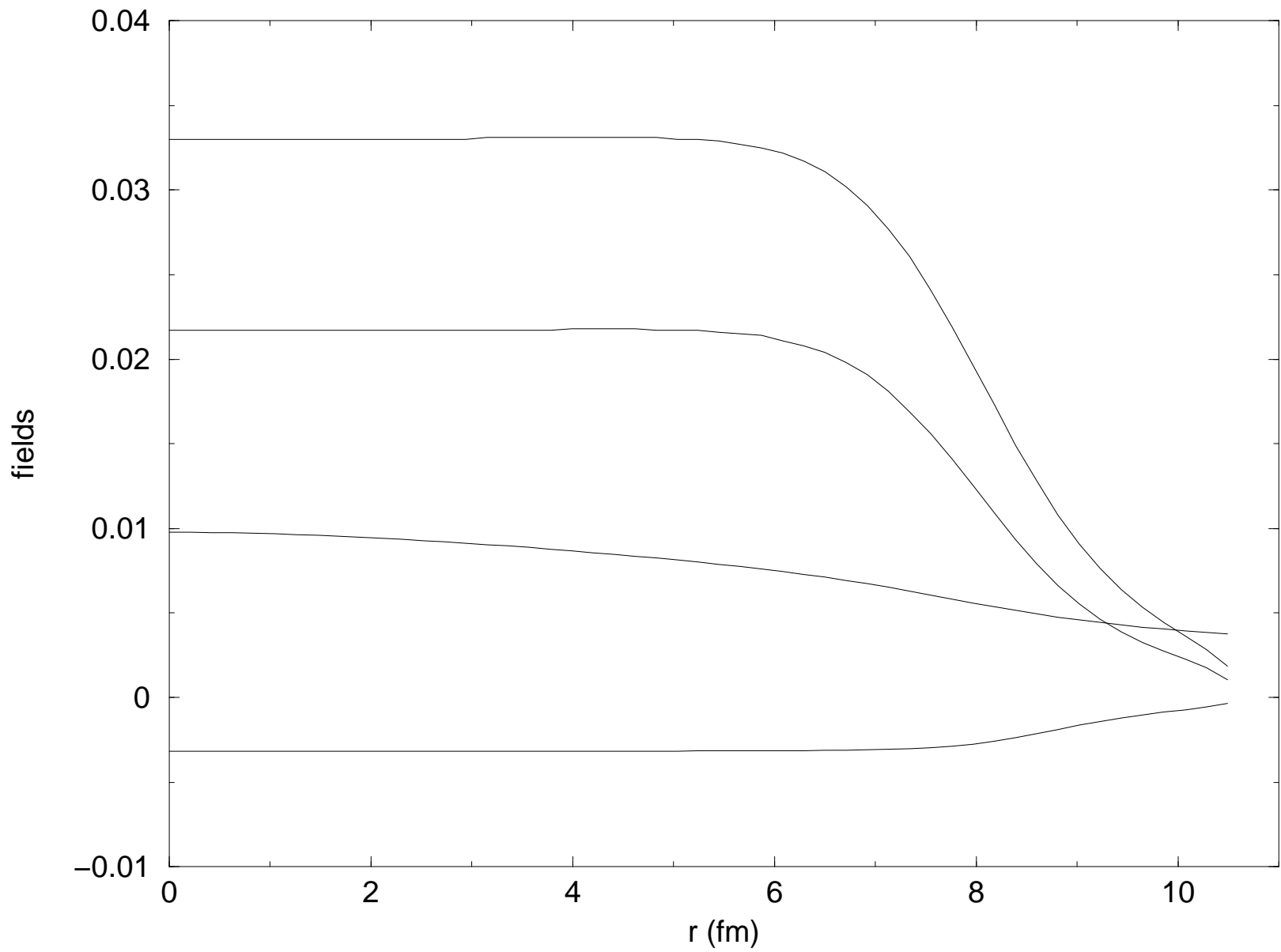
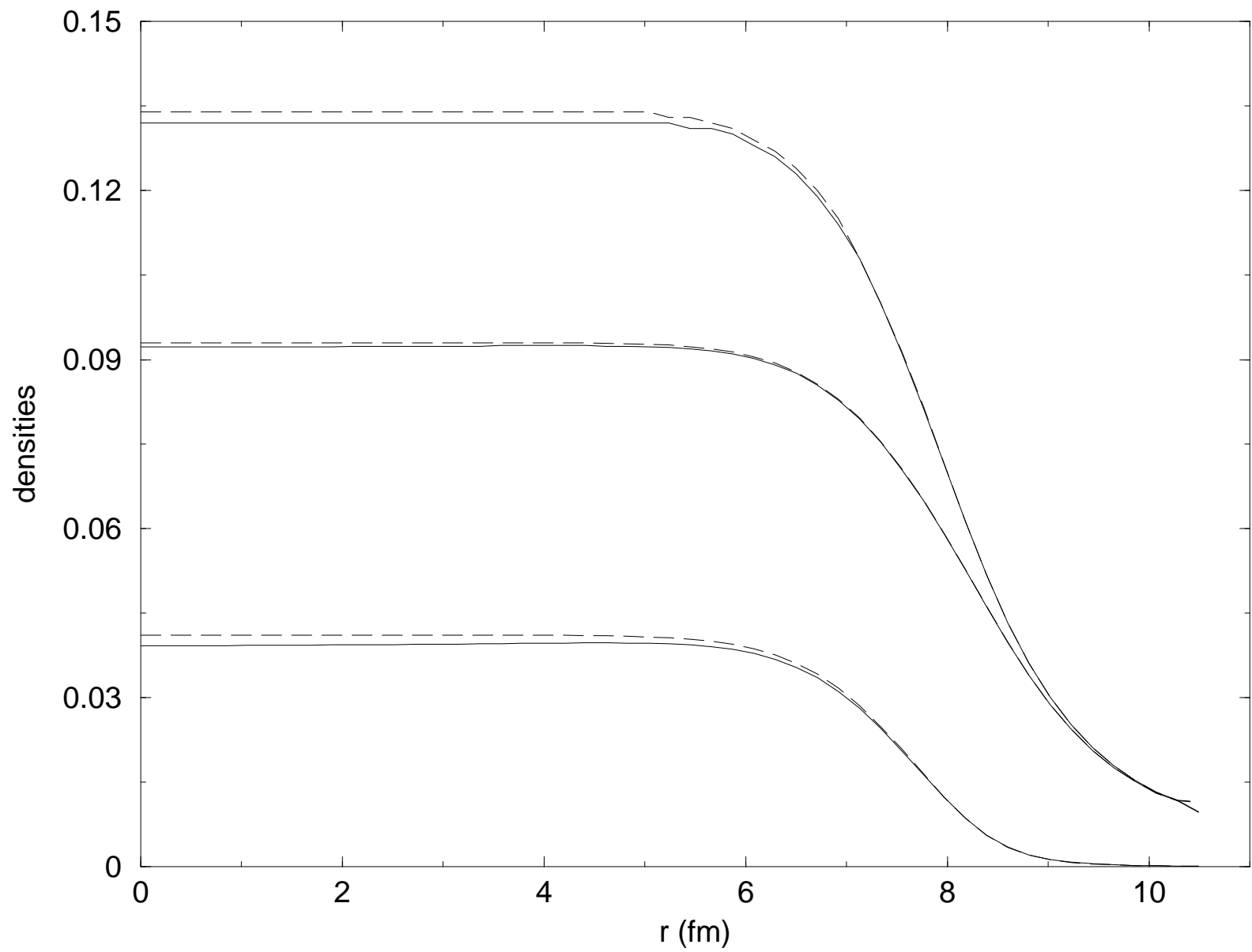
\end{document}